\documentclass[aps,prb,twocolumn,groupedaddress]{revtex4}
\usepackage{graphicx}
\usepackage{amsmath}
\graphicspath{{./}{./EPS/}}

\newcommand{\half}{\mbox{$\textstyle \frac{1}{2}$}}
\newcommand{\ket}[1]{\left | \, #1 \right \rangle}
\newcommand{\bra}[1]{\left \langle #1 \, \right |}

\begin{document}

\title{Fractional-quantum-Hall edge electrons and Fermi
statistics}

\author{U.~Z\"ulicke}
\affiliation{Institut f\"ur Theoretische Festk\"orperphysik,
Universit\"at Karlsruhe, D-76128 Karlsruhe, Germany}

\author{J.~J.~Palacios}
\affiliation{Departamento de F\'{\i}sica Aplicada, Universidad de
Alicante, San Vicente del Raspeig, Alicante 03690, Spain}

\author{A.~H.~MacDonald}
\affiliation{Department of Physics, The University of Texas at
Austin, Austin, TX 78712}

\date{\today}

\begin{abstract}

We address the quantum statistics of electrons created in the
low-energy edge-state Hilbert space sector of incompressible
fractional quantum Hall states, considering the possibility that
they may not satisfy Fermi statistics.  We argue that this
property is not {\it a priori\/} obvious, and present numerical
evidence based on finite-size exact-diagonalization calculations 
that it does not hold in general.  We discuss different possible
forms for the expression for the electron creation operator in
terms of edge boson fields and show that none are consistent with
our numerical results on finite-size $\nu=2/5$ states with
short-range electron-electron interactions. Finally, we discuss
the current body of experimental results on tunneling into
quantum Hall edges in the context of this result.

\end{abstract}


\maketitle

\section{Introduction}

The quantum Hall (QH) effect~\cite{qhe-sg,ahmintro} occurs when
an incompressibility, {\it i.e.}, a discontinuity in the dependence
of chemical potential $\mu$ on density, occurs at a
two-dimensional (2D) electron system (ES) sheet density $n_0$
that is magnetic-field $B$ dependent.  The QH effect at integer
Landau-level filling factors $\nu = 2\pi\ell^2 n_0$ arises 
from the quantization of 2D ES kinetic energy and from the
macroscopic degeneracy of Landau-level states with a particular
kinetic energy. ($\ell=\sqrt{\hbar c/|e B|}$ is known as the
magnetic length.) The fractional QH effect (chemical potential
jumps at fractional values of $\nu$), on the other hand, does not
occur in a non-interacting electron system, and is due to
constraints on the correlations\cite{rbl:prl:83} that can be
achieved among electrons that have the same quantized kinetic
energy. A necessary consequence\cite{ahmintro} of magnetic-field
dependence in $n_0$ is the existence of states in the
chemical-potential gap that are localized at the edge of a
finite-size system and carry equilibrium current $I$ with the
property $dI/d\mu=\nu e/h$.\cite{rbl:prb:81} These edge-electron
systems are obviously one-dimensional and, since they carry an
equilibrium current, obviously chiral.\cite{froh:nuclB:91}
Microscopically,\cite{bih:prb:82} the edge of a non-interacting
electron system at integer filling factor $\nu=m$ is equivalent,
at low energies, to a one-dimensional electron system with $m$
flavors of fermions that can travel only in one direction, {\it
i.e.}, $m$ chiral fermion branches. It was argued some time
ago,\cite{ahm:prl:90,wen:prb:90} on the basis of trial
wavefunctions for finite-size systems, that the edges of
incompressible {\em fractional\/} QH states can also be described
microscopically as chiral one-dimensional electron fluids that
have, in general, an unequal number of inequivalent left-moving
and right-moving branches. In a series of beautiful papers based
on hydrodynamic and field-theoretic considerations,
Wen\cite{wen:prb:90,wen:int:92,wen:adv:95} proposed and developed
the idea that the properties of fractional QH edges could be
described using a generalization of the bosonization approach
that was developed earlier for conventional one-dimensional
electron
systems\cite{schott:pr:69,lut:prb:74,emery,voit:reprog:94} and
provides a simple description of their
characteristic\cite{fdmh:jpc:81} {\em Luttinger-liquid\/} power-law
correlation functions. In Wen's theory, edge excitations can be
described microscopically in terms of a number of chiral boson
fields, and the resulting edge system is a {\em chiral Luttinger
liquid\/} ($\chi$LL).  Finite-size numerical
calculations~\cite{jujo:prl:96} have been a useful tool in
verifying the fundamentally bosonic character of the
edge-excitation spectrum and in testing some experimental
predictions of $\chi$LL theory. Further experimental evidence in
support of some aspects of $\chi$LL theory is summarized below.
There are, however, difficulties in reconciling this effort to
capture generic aspects of the microscopic physics of fractional
QH edges with experimental observations. Early
work\cite{amc:prl:96} did appear to imply that the edge structure
of QH systems at the Laughlin series of filling factors, {\it
i.e.,} for $\nu=1/(2 p+1)$ with positive integer $p$, is rather
well-described in terms of a single-branch $\chi$LL characterized
by a power-law exponent $\alpha=1/\nu$, as expected on the basis
of $\chi$LL considerations. Recently, however, this universal
dependence of $\alpha$ on filling factor has been questioned both
theoretically\cite{gold:prl:01,gold:prb:01,jain:ssc:01,kun:prl:02}
and experimentally.\cite{hilke:prl:01} In addition, tunneling
density-of-states observations\cite{amc:prl:98} at hierarchical
filling factors, {\em i.e.}, $\nu=m/(2 m p\pm 1)$ with integer
$m>1$ are in apparent contradiction with the
predictions\cite{mpaf:prb:95a} of $\chi$LL theory.

Motivated by this stark experimental discrepancy, we re-examine
in this paper a key {\em ansatz\/} of $\chi$LL theory, which has
appeared to us to be non-obvious\cite{uz:prb:99} and concerns
the properties of the operator $\tilde\psi^\dagger$ obtained from
the full microscopic electron creation operator $\psi^\dagger$ by
{\em projecting\/} onto the low-energy sector of edge excitations:
\begin{equation}\label{project}
\tilde\psi^\dagger = {\mathcal P}\,\psi^\dagger\,{\mathcal P}
\quad ,
\end{equation}
where ${\mathcal P}=\sum_{\{\eta_i\}}\ket{\{\eta_i\}}\bra{\{
\eta_i\}}$ is the projection operator onto the Fock-space subset
of low-energy (edge) excitations $\ket{\{\eta_i\}}$, discussed at
greater length below. In applying $\chi$LL theory to evaluate
electronic correlation functions, the representation of $\tilde
\psi$ in terms of bosonic edge-density fluctuations is a key
ingredient.  Such bosonization identities can be derived
constructively\cite{shankar:pol:95,hammer:unpub,vondelft} for a
conventional one-dimensional electron
system.\cite{voit:reprog:94} Their generalization to the
fractional-QH case,\cite{wen:int:92,mpaf:int:94} in which
there is no adiabatic connection to non-interacting electron
states, must however be based on appealing but heuristic physical
arguments that, ultimately, have to be verified by experiment, or
by numerical calculations. In our view it is not clear beyond any
doubt that the bosonized forms of $\tilde\psi$ that have been
used in $\chi$LL theory to evaluate electronic correlation
functions and predict observables, like the power-law exponents
in tunneling current-voltage characteristics, are always correct.
In particular, an important guiding principle that has been used
to limit possible bosonized expressions for electron field
operators in $\chi$LL theory is the seemingly obvious requirement
that they satisfy Fermi statistics. In this paper, we use
finite-size exact diagonalization studies of a
short-range-interaction model to directly test the
Fermi-statistics {\it ansatz}. The subset of Fock-space states that
represent edge excitations of this model can, for the most part,
be identified convincingly. {\em We demonstrate by explicit calculation
of some anticommutator matrix elements that
electron creation operators projected onto this low-energy Fock
space do not satisfy Fermi anticommutation rules.} In
Section~\ref{num} of this paper, the numerical calculations that
support this claim are described in detail. Section~\ref{boscons}
discusses the problem of understanding the properties of the
projected electron creation operator and of finding a useful
expression for it in terms of edge boson fields, in light of our
numerical finding. We conclude in Section~\ref{concl} with a
brief summary.  A preliminary report on this work was presented
earlier.\cite{juanjo:baps:00}

\section{Numerical test of the Fermi statistics
\textbf{\textit{ansatz}}}\label{num}

The second-quantized operator $\psi^\dagger(x,y)$ [$\psi(x,y)$]
that creates [annihilates] 2D electrons in the lowest Landau
level obeys anticommutation relations that encode the fundamental
antisymmetry condition satisfied by many-fermion wavefunctions: 
\begin{equation}\label{commut}
\left\{\psi^\dagger(x,y), \psi^\dagger(x^\prime, y^\prime)\right
\} = 0\quad .
\end{equation}
The question we address in this section is whether the
anticommutation relations are still satisfied after {\em
projection\/} onto the low-energy (long-wavelength) sectors of
Fock-space that represent edge excitations of particular
incompressible states. Because of the projection,
Eq.~(\ref{commut}) does not mathematically guarantee the relation
\begin{equation}\label{commrel}
\tilde\psi^\dagger(x,y)\tilde\psi^\dagger(x^\prime,y^\prime)=
-\tilde\psi^\dagger(x^\prime,y^\prime)\tilde\psi^\dagger(x,y)
\quad .
\end{equation}
A physical argument along the following lines does, however,
appear plausible.  It is possible to add two different electrons
to the system at low energies that are localized at different
positions along the edge via processes that can be represented by
the projected creation operator. The edge-electron positions can
then be adiabatically interchanged, ending up with an equivalent
many-fermion state that must differ only by a sign from the
original state. If the many-particle state can be represented, at
all intermediate relative positions, by two projected creation
operators acting on the starting state, it seems hard to escape
the conclusion that these operators must satisfy Fermi
statistics. However, the correlations that establish the bulk gap
could be disturbed when the two electrons are in close proximity.
It is therefore difficult to exclude the possibility that this
argument breaks down at the crossing point in the exchange path.
Related arguments can be advanced in which the exchange paths
involve particle creation at different times, but do not appear
to us to be conclusive. Our inability to settle this point on the
basis of simple general arguments has motivated the numerical 
calculations we now explain.

The identification of the set of states as edge excitation states
of a particular incompressible state in a finite-size
many-fermion spectrum is both a challenge and an important source
of uncertainty for the conclusions we reach. For $\nu=1/(2p+1)$,
the identification is accomplished\cite{jujo:prl:96} by appealing
to Laughlin\cite{rbl:prl:83,ahm:dbm:prb:85} to conclude that the
low-energy edge-excitation states appear at angular momenta above
$L_0=(2p+1)N(N-1)/2$, and that the dimension of subspaces at
fixed angular momentum is related to their excess momentum by
counting the number of modes in a chiral boson Hilbert
space;\cite{ahm:braz:96} {\it i.e.}, one state with angular
momentum 1, two with angular momentum 2, three with angular
momentum 3, five with angular momentum 4, {\it et cetera}. Previous
numerical work\cite{jujo:prl:96} has verified $\chi$LL
predictions for $\nu=1/(2p+1)$ edges, but not at filling factors
within the range $1/3 < \nu < 1$, where experiment and theory
appear to be at odds.

Incompressibilities at many other values of the filling factor,
{\it e.g.}, for $\nu=m/(2 m p\pm 1)$ with positive integer $m$,
have been explained using various hierarchical
schemes\cite{fdmh:prl:83,bih:prl:84} and using composite-fermion
theory.\cite{jain:prl:89} Composite-fermion theory and
variational wavefunctions based on hierarchy-theory ideas make
identical predictions\cite{read:prl:90} for the values of total
angular momentum $L_0$ at which
{\em maximum-density\/}\cite{ahm:braz:96} incompressible states
(those with edge subsystems in their ground states) appear and
for the number and chirality of the boson branches in their
edge-excitation spectra. For practical reasons explained more
fully below, we limit our study to QH systems with two branches
of edge excitations that have the same {\em chirality}, {\it i.e.},
that propagate along the edge in the same direction and have, for
our circular droplets, excess angular momenta of the same sign.
This is the case for QH systems at filling factors $\nu=2/(4 p+1)
$ where the edge spectrum is expected to be that of two boson
modes that share the same chirality.  

The angular-momentum values at which maximum-density states occur
depend on the number of electrons $N$, and the number of
particles to be transferred from the ground to the upper
composite-fermion Landau level $N_{qp}$. This notation is chosen
to suggest the analogous hierarchy-picture description of the
same states, which makes identical predictions for the set of
angular-momentum values at which maximum-density states occur.
In the composite-fermion language, the number of particles in the
lower composite-fermion Landau level is $N-N_{qp}$, and {\em
maximum-density\/} finite-size $\nu=2/(4p + 1)$ states appear at
$L_0=(N_{qp}-1)*(N_{qp}-2)/2+(N-N_{qp})(N-N_{qp}-1)/2+2pN (N-1)$,
where the last term comes from the Jastrow factor in Jain's
variational wavefunctions and the first term reflects the reduced
angular momentum of higher Landau-level states.\cite{devjain}
Changing $N_{\text{qp}}$ and/or $N$ is analogous to a {\em
topological\/} or {\em zero-mode\/} excitation in a finite-size
one-dimensional electron system.\cite{fdmh:jpc:81} We denote the
N-electron state with $N_{\text{qp}}$ quasiparticles and an
edge-density sub-system in its ground state by $\ket{N,
N_{\text{qp}},0}$, and its total momentum by $L_{\text{0}}(N,
N_{\text{qp}})$. The following relation can be derived from the
above composite-fermion expression or from hierarchy theory:
\begin{equation}\label{hierpred}
L_0(N, N_{\text{qp}}) = (2p+1)\frac{N(N-1)}{2} + N_{\text{qp}}
(N_{\text{qp}}-1) - N\, N_{\text{qp}}\, .
\end{equation}

\begin{figure*}
\includegraphics[width=3.3in]{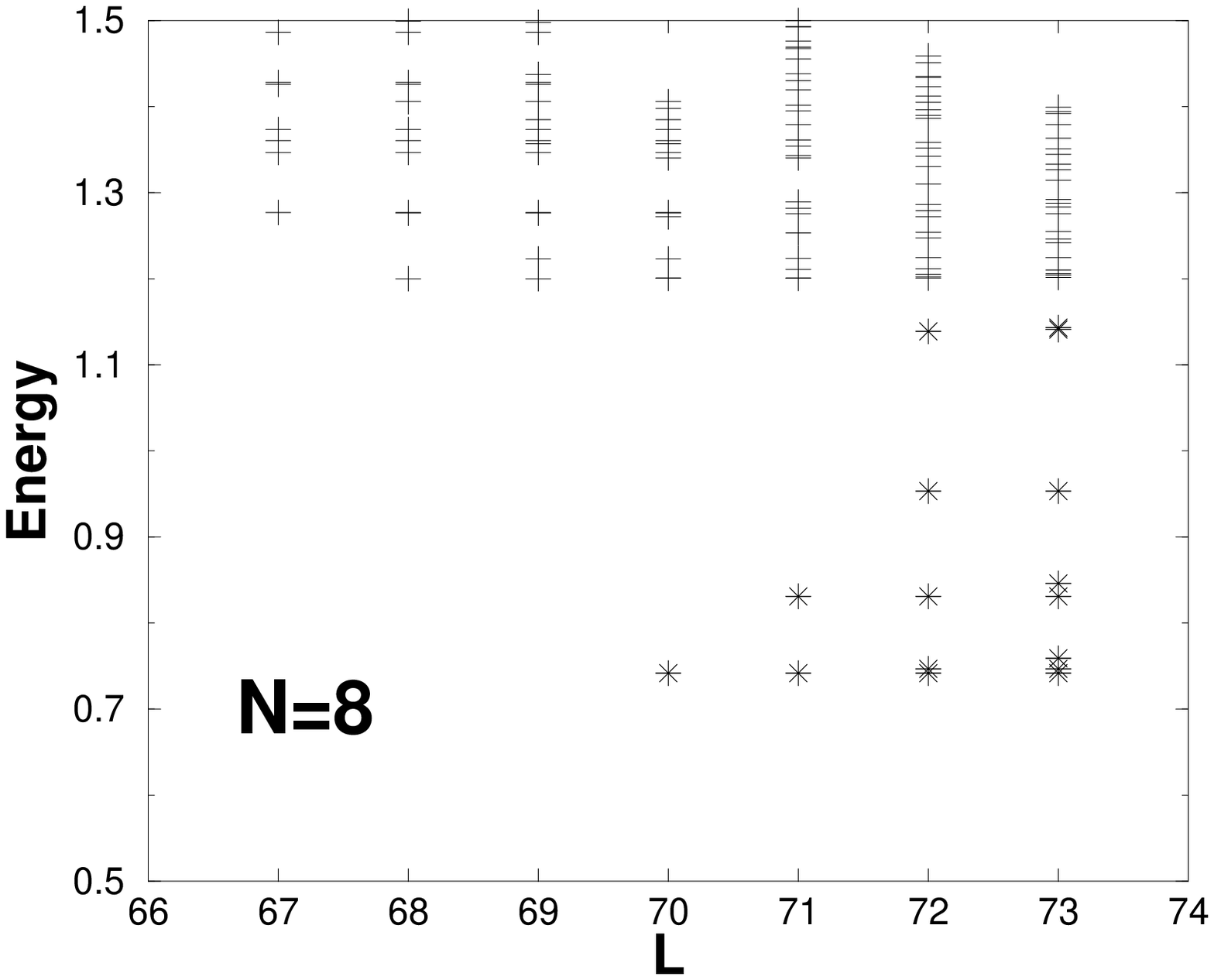}\hfill
\includegraphics[width=3.3in]{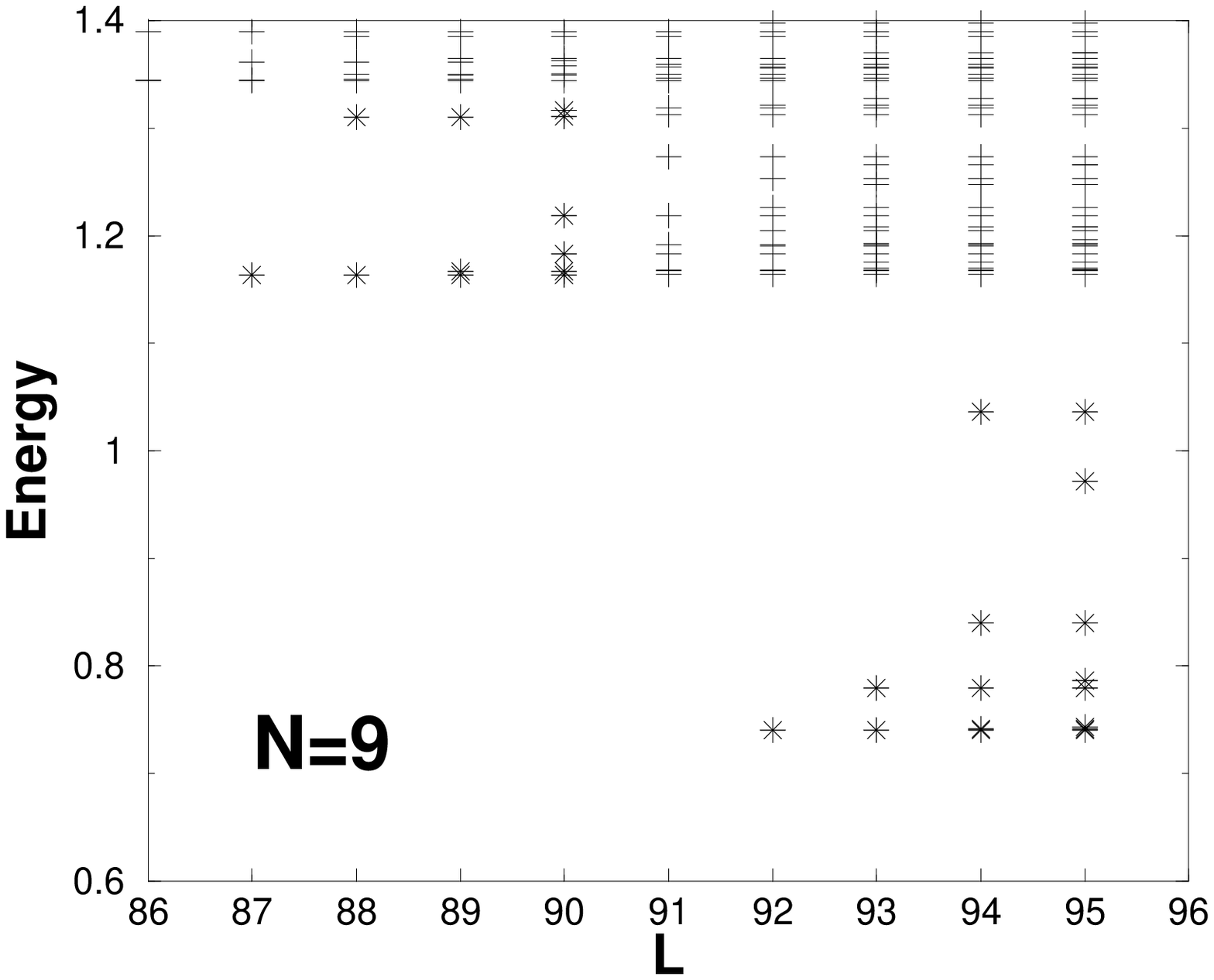}
\caption{Low-energy portion of the exact finite-size spectrum for
$N=8$ and $N=9$ particles with short-range interactions. The
non-degenerate states at $L=70$ ($N=8$), $L=87$ ($N=9$), and
$L=92$ ($N=9$) can be identified as maximum-density ground states
of a $\nu=2/5$ finite-size QH droplet. Their energy is separated
by a large gap from a continuum of bulk-excited states. States at
higher angular momenta with energies below that gap (indicated by
asterisks) can be unambiguously identified as edge excitations.
Note that the multiplicity of these states at $L=71$ and $L=72$
in the left panel and $L=88$, $93$, and $94$ in the right panel
is exactly as expected for a two-branch chiral boson system. For
$L\ge 73$ in the left panel and $L\ge 89$ and $L\ge 95$ in the
right panel, `edge' states exist that have energies as high as
bulk excitations. In the calculations described below, we try to
select anticommutator matrix elements that do not require us to
utilize Fock-space sectors with missing edge states. Cases for
which there are missing states are explicitly identified in our
discussion. The energies plotted here do not include the
contribution, proportional to total angular momentum, from our
model's parabolic confinement potential which lifts the energies
of states with larger $L$. The ground state is determined by the
strength of this confinement potential. In the thermodynamic
limit, it will be one of the $\nu=2/5$ maximum-density-droplet
states when the chemical potential lies in the $\nu=2/5$ gap.
\label{spectra}}
\end{figure*}

\begin{table}
\caption{Total angular momenta $L_0(N,N_{\text{qp}})$ for the
ground state of compact fractional-QH systems as predicted from 
Eq.~(\ref{hierpred}) for $p=1$. These states can be regarded as
$N$-electron states in the lowest Landau level that consist of a
QH droplet at filling factor $1/3$ supporting a compact daughter
droplet of $N_{\text{qp}}$ quasiparticles. In an equivalent
description, these are $N$-composite-fermion states with lowest
and first-excited composite-fermion Landau levels having
occupation $N-N_{\text{qp}}$ and $N_{\text{qp}}$, respectively. 
Numbers in bold type indicate that the corresponding ground state
in our numerical spectra was positively identified as a
finite-size $\nu=2/5$ QH state.\protect\footnote{We do not
subscribe to the different interpretation of some of these states
given by \citet{capi:prb:98} based on density profiles.} The
ground states whose edge-excitation sectors are explicitly used
here to test the Fermi statistics {\it ansatz\/} are indicated by
asterisks.
\label{momenta}}
\begin{ruledtabular}
\begin{tabular}{cc|rrrrrr}
 & $N$ & 5 &         6 &       7 &      8 &        9 &     10 \\ 
 $N_{\text{qp}}$  &  & \\ \hline
0 & &    30 & $^\ast$45 &      63 &     84 &      108 &    135 \\
1 & &{\bf 25}&   {\bf 39} & {\bf 56} & {\bf 76} &  {\bf 99} & 125 
\\
2 & &{\bf 22}&{\bf $^\ast$35}&{\bf $^\ast$51}& {\bf $^\ast$70}&
{\bf $^\ast$92} & {\bf 117} \\
3 & &    & {\bf 33} &      48 & {\bf $^\ast$66} & {\bf $^\ast$87}&
{\bf 111} \\
4 & &         &         &         &      64 &  84  &    107 \\
5 & &         &         &         &         &         &    105 
\end{tabular}
\end{ruledtabular}
\end{table}
For the case $p=1$, Table~\ref{momenta} shows $L_0(N,
N_{\text{qp}})$ for these finite-size {\em maximum-density\/} states.
Finite-size numerical spectra for small $N$ exhibit
non-degenerate ground states at these values of $L$, and a
low-energy excitation spectrum at small excess $L$ that
corresponds to the expected chiral two-branch edge when $N_{qp}$
is close enough to $\approx N/2$. (A chiral two-branch edge has
two low-energy states with excess angular momentum equal to 1,
five with excess angular momentum 2, ten with excess momentum 3,
{\it et cetera}. As an example, Figure~\ref{spectra} shows spectra
obtained for $N=8$ and $9$ particles.) We define the edge-state
Fock-space projection ${\mathcal P}$ by retaining for each $N$
and $N_{\text{qp}}$ only these states.

The low-energy Fock space is then the direct sum of the
low-energy Hilbert spaces for different particle numbers $N$;
each of these is in turn the direct sum of orthogonal subspaces
labeled by different values of $N_{\text{qp}}$, the number of
composite fermions in the first excited Landau level.  Our
understanding of $\chi$LL theory is that it attempts to describe
the physics of QH systems projected onto this subspace. It will
prove useful to write the projected electron creation operator as
the sum of separate contributions labeled by the change in the
number of quasiparticles that accompanies a particle addition. We
therefore write $\tilde\psi^\dagger=\sum_n\tilde\psi^\dagger_n$
where the operator $\tilde\psi^\dagger_n$ changes $N_{\text{qp}}$
by $n$. In other words, an electron can be added to an
$N$-particle, $N_{\text{qp}}$-quasiparticle finite-size state in
many different ways, distinguished both by the resulting edge
disturbance and by the number of quasiparticles in the resulting
$(N+1)$-particle state. If the corresponding projected electron
creation operators $\tilde\psi^\dagger_n$ satisfy Fermi
statistics, their anticommutator $\{\tilde\psi_{n^\prime}^\dagger
,\,\tilde\psi_{n}^\dagger\}$ will vanish identically. We check
for Fermi statistics by evaluating anticommutator matrix elements
that involve the lowest possible excess momenta and are therefore
likely to have the smallest finite-size effects. In particular,
we define 
\begin{widetext}
\begin{subequations}\label{define}
\begin{eqnarray}
A_{n,n^\prime}(N,N_{\text{qp}})&=&\bra{N+2,N_{\text{qp}}+n+
n^\prime,0} c^\dagger_{K_2}{\mathcal P}_{N+1,N_{\text{qp}}+n}\,
c^\dagger_{K_1}\ket{N,N_{\text{qp}},0}\quad , \\
B_{n,n^\prime}(N,N_{\text{qp}})&=&\bra{N+2,N_{\text{qp}}+n+
n^\prime,0}c^\dagger_{K_1}{\mathcal P}_{N+1,N_{\text{qp}}+
n^\prime}\, c^\dagger_{K_2}\ket{N, N_{\text{qp}},0}\quad ,
\end{eqnarray}
\end{subequations}
with $K_1=K_{\text{F}}^{(n)}(N,N_{\text{qp}})$ and $K_2=
K_{\text{F}}^{(n^\prime)}(N+1,N_{\text{qp}}+n)$ where
$K_{\text{F}}^{(n)}(N,N_{\text{qp}}):= L_0(N+1,N_{\text{qp}}+n)-
L_0(N,N_{\text{qp}})$. The projector ${\mathcal
P}_{N,N_{\text{qp}}}$ projects onto the subspace of the
$N$-particle low-energy subspace having $N_{\text{qp}}$
quasiparticles ({\it i.e.}, composite fermions in their first
excited Landau level). It is then obvious that
\begin{subequations}
\begin{eqnarray*}
A_{n,n^\prime}(N,N_{\text{qp}})&=&\bra{N+2,N_{\text{qp}}+n+
n^\prime,0}c^\dagger_{K_2}\ket{N+1,N_{\text{qp}}+n,0}\bra{N+1,
N_{\text{qp}}+ n,0}c^\dagger_{K_1}\ket{N,N_{\text{qp}},0}\quad ,
\\ B_{n,n^\prime}(N,N_{\text{qp}})&=&\sum_{\{\eta\}}\bra{N+2,
N_{\text{qp}}+n+n^\prime,0}c^\dagger_{K_1}\ket{N+1,N_{\text{qp}}+
n^\prime,\{\eta\}}\bra{N+1,N_{\text{qp}}+n^\prime,\{\eta\}}
c^\dagger_{K_2}\ket{N, N_{\text{qp}},0}\, ,
\end{eqnarray*}
\end{subequations}
\end{widetext}
where $\ket{N+1,N_{\text{qp}}+n^\prime,\{\eta\}}$ are edge
excitations in the $(N+1)$-particle, $(N_{\text{qp}}+n^\prime)
$-quasiparticle system for {\em excess\/} total momentum
\begin{eqnarray}\label{deltafermi}
\Delta K_{\text{F}}^{(n,n^\prime)}&=& K_{\text{F}}(N+1,
N_{\text{qp}}+n,n^\prime)-K_{\text{F}}(N,N_{\text{qp}},n^\prime)
\, , \nonumber \\
&=& 2p+1+2n n^\prime - n - n^\prime\quad .
\end{eqnarray}
Fermi statistics is satisfied if the ratio $B/A$ equals $-1$.

\begin{table}
\caption{Results for the ratio between the matrix elements $A=
\langle L_0^{(N+2)}|c^\dagger_{K_2}{\mathcal P}\, c^\dagger_{K_1}
|L_0^{(N)}\rangle$ and $B=\langle L_0^{(N+2)}|c^\dagger_{K_1}
{\mathcal P}\, c^\dagger_{K_2}|L_0^{(N)}\rangle$. This ratio
should always equal $-1$ if the edge--electron operator satisfies
Fermi statistics. We have chosen values $K_1$ and $K_2$ such that
$A = \langle L_0^{(N+2)}|c^\dagger_{K_2}|L_0^{(N+1)}\rangle
\langle L_0^{(N+1)}|c^\dagger_{K_1}|L_0^{(N)}\rangle$ involves
matrix elements between {\em maximum-density\/} states only. When the
order of edge--electron operators is reversed, $B=\sum_r \langle
L_0^{(N+2)}|c^\dagger_{K_1}|{\tilde L}_r^{(N+1)}\rangle\langle
{\tilde L}_r^{(N+1)}|c^\dagger_{K_2}|L_0^{(N)}\rangle$, with the
sum ranging over matrix elements between edge states $|{\tilde
L}_r^{(N+1)}\rangle$ with excess angular momentum $\Delta K =
L_0^{(N)}+K_2 - {\tilde L}_0^{(N+1)}$. The first block
corresponds to $B_{0,0}/A_{0,0}$, the second to $B_{0,1}/A_{0,1}$
and $B_{1,0}/A_{1,0}$, and the third to $B_{1,1}/A_{1,1}$.
\label{restab}}
\begin{ruledtabular}
\begin{tabular}{ccccccccc}
$N$ & $L_0^{(N)}$ & $L_0^{(N+1)}$ & $L_0^{(N+2)}$ & $K_1$ & $K_2$
& ${\tilde L}_0^{(N+1)}$ & $\Delta K$ & $B/A$ \\ \hline
5 & 30 & 45 & 63 & 15 & 18 & 45 & 3 & $-1$\footnote{The
edge--electron operator for any finite--size Laughlin state
satisfies Fermi statistics perfectly.} \\
5 & 22 & 35 & 51 & 13 & 16 & 35 & 3 & $-0.78$\footnote{Incomplete
set of edge states.} \\
6 & 35 & 51 & 70 & 16 & 19 & 51 & 3 & $-0.99$\footnotemark[2] \\
7 & 51 & 70 & 92 & 19 & 22 & 70 & 3 & $-1.02$\footnote{Nearly
complete set of edge states.}\\
8 & 70 & 92 & 117 & 22 & 25 & 92 & 3 & $-1.01$\footnotemark[3]\\
8 & 66 & 87 & 111 & 21 & 24 & 87 & 3 & $-1.05$\footnotemark[2] \\
\hline
6 & 39 & 56 & 70 & 17 & 14 & 51 & 2 & $-1.12$\footnotemark[3] \\
7 & 56 & 76 & 92 & 20 & 16 & 70 & 2 & $-0.95$ \\
7 & 51 & 66 & 87 & 15 & 21 & 70 & 2 & $-1.24$ \\
7 & 51 & 70 & 87 & 19 & 17 & 66 & 2 & $-1.32$\footnotemark[2] \\
8 & 70 & 92 & 111 & 22 & 19 & 87 & 2 & $-1.14$\footnotemark[2] \\
8 & 70 & 87 & 111 & 17 & 24 & 92 & 2 & $-1.11$ \\
8 & 76 & 99 & 117 & 23 & 18 & 92 & 2 & $-0.88$ \\
\hline
6 & 39 & 51 & 66 & 12 & 15 & 51 & 3 & $-1.44$\footnotemark[2] \\
7 & 56 & 70 & 87 & 14 & 17 & 70 & 3 & $-1.26$\footnotemark[3] \\
8 & 76 & 92 & 111 & 16 & 19 & 92 & 3 & $-1.20$\footnotemark[3] \\
\end{tabular}
\end{ruledtabular}
\end{table}

Our numerical results are summarized in Table~\ref{restab}. We
start by considering a QH droplet at filling factor $1/3$, {\it
i.e.}, states with $N_{qp}=0$. (See the first row in
Table~\ref{restab}.) Although only one case is presented there,
it turns out that for {\em any\/} finite-size system with
short-range interactions, the electron operator projected onto
the subspace of low-energy excitations above the ground state at
$\nu=1/3$ satisfies Fermi statistics {\em exactly}. The Fermi
statistics of edge excitations is a property of Laughlin
many-body wavefunctions,\cite{rbl:prl:83} which are exact in the
case of the hard-core-interaction model at $\nu=1/(2p+1)
$.\cite{jjp:note} It is apparent from the remaining entries in
Table~\ref{restab} that this property of hard-core-model
single-branch QH edges does not, in general, hold in the $\nu=2/5
$ two-branch case. We have tested edge-excitation subspaces for
systems with $N=6,7,8$, and $9$ electrons and find that the ratio
$B/A$ that `measures' Fermi statistics often differs
substantially from $-1$. Instead, a strong dependence of $B/A$ on
the change in $N_{qp}$ that accompanies particle addition is
apparent from our data. For example, consider the sequences
$L^{(N)}$--$L^{(N+1)}$--$L^{(N+2)}$ where particles are added
without adding hierarchy quasiparticles. In composite-fermion
language, this corresponds to adding electrons to the lowest
composite-fermion Landau level only. The finite sequences in this
class that we have studied, {\it e.g.}, 35--51--70, 51--70--92,
and 66--87--111, are gathered in the first block of
Table~\ref{restab}. Although violations of Fermi statistics are
numerically unambiguous, they are large only for those cases
where the edge sector is seriously incomplete (only 5 out of 10
edge states could be identified in the L=38 sector for the
sequence 22--35--51). To some extent, this result is not
surprising since the addition of an electron to the lowest
composite-fermion Landau level is expected to involve excitations
of this level only, as in the $\nu=1/3$ case. It seems plausible
that the results should be similar. The deviation from Fermi
statistics is suspicious, however, when contrasted with the
perfect accuracy seen for the $\nu=1/3$ QH droplet. When an
anticommutator involving $\tilde\psi^\dagger_1$ (increasing the
number of higher-Landau-level composite fermions by one) is
considered, as in the sequences 39--51--66, 56--70--87, and
76--92-111 (see the third block in Table~\ref{restab}), the
deviations from Fermi statistics are larger. Note that some of
the {\em same\/} finite-size edge states are used here and in the
$\tilde\psi^\dagger_0$ case for which Fermi statistics is more
closely approximated. It is important to recognize, however, that
the edge sectors used in all these sequences are not complete
even for the larger systems considered. The incompleteness is
related to the fact that there are only two electrons in the
first excited composite-fermion Landau level which implies that
the {\em daughter\/} droplets are substantially smaller than the
parent droplets, exacerbating finite-size difficulties. The
strongest evidence that Fermi-statistics relations are not
satisfied comes from the remaining sequences in
Table~\ref{restab}, for which the numerical edge-state sector is
usually complete. There is no unique way of estimating a
thermodynamic limit for $B/A$ from our data. A consistent scheme
would have to keep $N-2 N_{\text{qp}}\ll N$ as $N\rightarrow
\infty$, in order to maintain parent and daughter fluids that are
similar in size. The small system sizes tractable using current
numerical methods render such an extrapolation impossible.
Examining the trends in Table~\ref{restab}, however, we are
reasonably confident that differences $|1+B/A|>0.05$ are
significant and not merely due to finite-size effects. {\em We
note that in the case of conventional one-dimensional electron
systems, a corresponding calculation would always result in exact
conformation with Fermi statistics; there are no finite-size
corrections.} The $\nu=1/3$ hard-core-model case also produces
results in agreement with Fermi statistics without finite-size
corrections. In contrast, our calculations exhibit large
deviations from Fermi statistics for many sequences where the
edge states expected for a two-branch boson system are clearly
resolved in the spectrum, not only for those with an incomplete
edge sector. This is the case for 51--66--87 for example.
Furthermore, when edge states exist that have energies above the
gap for bulk excitations and can no longer be clearly identified
(this is the case, e.g., for 8-particle states at total momentum
73), the value of $B/A$ turns out to be determined almost
entirely by contributions from edge states with energies below
the gap. Inclusion of any number of states (bulk or edge) above
the gap energy changes $B/A$ only by a few percent. Finally,
deviations from Fermi statistics do not seem to diminish with
increasing particle number. On the contrary, the more
unambiguously identified edge-state sectors at larger $N$ yield
values of $B/A$ that differ consistently from $-1$, especially
for anticommutators involving the component of the creation
operator that increases the number of daughter quasiparticles,
$\tilde\psi_1^\dagger$.

\section{Properties of Candidate Bosonized Electron Operators}
\label{boscons}

Our numerical results clearly support a multi-branch chiral-boson
form for the excitation spectrum of a fractional-QH edge, but
raise new questions about the representation of projected
electron creation operators in terms of these boson fields. This
issue is discussed in the following section. We start by
carefully examining the arguments that have been made in $\chi$LL
theory to obtain bosonization identities. Some of these heuristic
arguments must be ruled out if the edge projection of the
electron operator does indeed not satisfy Fermi statistics.
Alternative proposals are discussed, but we have not been able to
find a simple form that is consistent with our numerical results,
suggesting that the true expression may not be universal.

In a conventional one-dimensional system,\cite{voit:reprog:94}
the low-energy projection of the electron operator
$\psi^\dagger_{\text{1D}}$ is expressed as the sum $\tilde
\psi_{\text{1D}}^\dagger=\tilde\psi_{\text{R}}^\dagger+\tilde
\psi_{\text{L}}^\dagger$ of right-moving and left-moving chiral
fermion contributions $\tilde\psi_{\text{R,L}}^\dagger$. For
these chiral fermion operators, an identity relating them to the
bosonic charge fluctuations of an interacting system can be 
derived rigorously.\cite{shankar:pol:95,hammer:unpub,vondelft}
Fractional-QH edges {\em do\/} appear to be realizations of chiral
one-dimensional systems, as indicated by the multiplicity of
low-lying many-electron states in our numerically obtained
spectra. Given this observation, one is tempted to push the
analogy further and search for a bosonization identity for the
projected edge-electron operators $\tilde\psi_n^\dagger$.
Assuming that it is {\em local\/} in the angular coordinate
$\theta$ along the edge, it should read
\begin{equation}\label{simpbos}
\tilde\psi^\dagger_n(\theta)=\sqrt{z}\,e^{-i\theta
K_{\text{F}}^{(n)}(\hat N, \hat N_{\text{qp}})} e^{-i
\phi_n^\dagger(\theta)}e^{-i\phi_n(\theta)}\,
{\mathcal U}_n^\dagger \quad ,
\end{equation}
where $z$ denotes a normalization constant. The `Klein factor'
${\mathcal U}_n^\dagger$ is a ladder operator that connects
many-particle ground states, ${\mathcal U}_n^\dagger\ket{N,
N_{\text{qp}},0}=\ket{N+1,N_{\text{qp}}+n,0}$ and commutes with
bosonic edge-density operators. The correct commutation relations
for operators with {\em different\/} $n$, whatever they are, have to
be encoded in these factors. The chiral phase field $\phi_n(
\theta)$ is a superposition of edge-density fluctuations. For
$\nu=2/(4p+1)$, the following decomposition of the phase field in
terms of eigenmodes is always possible:
\begin{equation}\label{decomp}
\phi_n(\theta) = \frac{1}{\sqrt{\nu}}\,\phi^{\text{(c)}}(\theta)
+\xi_n\,\phi^{\text{(n)}}(\theta)\quad ,
\end{equation}
were $\phi^{\text{(c)}}$ is the phase field of the charged {\em
edge-magneto\-plasmon\/} mode which corresponds to fluctuations in
the total edge-charge density, and $\phi^{\text{(n)}}$ is its
orthogonal complement, the so-called neutral mode. The prefactor
$1/\sqrt{\nu}$ of the charged mode in Eq.~(\ref{decomp}) is
mandated by the fact that the addition of an electron necessarily
increases total electric charge by unity. Additional assumptions
are necessary, however, to fix the values of $\xi_n$.

Within $\chi$LL theory,\cite{wen:int:92} the operators $\tilde
\psi_0$ and $\tilde\psi_{2p+1}$ are believed to be special in
that they create electrons localized at the putative `outer' and
`inner' edges which are the boundaries of the outer parent and
inner daughter QH droplets with filling factors $\nu_{\text{o}}=1
/(2p+1)$ and $\nu_{\text{i}}=1/[(2p+1)(4p+1)]$, respectively,
that comprise the $\nu=2/(4p+1)$ QH state. The density
fluctuations at the `outer' and `inner' edges are given by
$(\nu_{\text{o}}/L)\,\partial_\theta\phi_0$ and $(\nu_{\text{i}}/
L)\,\partial_\theta\phi_{2p+1}$, respectively, and the definition
of charged and neutral modes implies that 
\begin{subequations}
\begin{eqnarray}
\phi^{\text{(c)}} &=& \frac{1}{\sqrt{\nu}}\left(\nu_{\text{i}}\,
\phi_{2p+1} + \nu_{\text{o}}\,\phi_0\right) \quad , \\
\phi^{\text{(n)}} &=& \sqrt{\frac{\nu_{\text{o}}\nu_{\text{i}}}
{\nu}}\left( \phi_{2p+1} - \phi_0 \right)\quad .
\end{eqnarray}
\end{subequations}
The addition of electrons to the edge with concomitant change of
$2p+1-n$ flux quanta is viewed as adding the electron to the
`outer' edge and transferring {\em at the same location\/} $n$
fractionally charged quasiparticles from the outer QH droplet to
the inner one. This suggests the relation $\tilde\psi_n^\dagger
\sim\tilde\psi_0^\dagger\,\exp\{i n\nu_{\text{o}}(\phi_0 -
\phi_{2p+1})\}$ which is equivalent to
\begin{equation}\label{allowed}
\xi_n = \frac{1}{\sqrt{2}}\,(2 n - 1)\quad .
\end{equation}

The chain of arguments leading to Eq.~(\ref{allowed}) involves
several assumptions that are not obviously satisfied. For
example, it is not clear why changes in $N_{qp}$ that accompany 
electron addition have to occur by transferring localized
quasiparticles from an `inner' edge to an `outer' one. Relaxing
this condition would lift the restriction expressed by
Eq.~(\ref{allowed}). In fact, two of the present authors
suggested the different choice $\xi_n=0$ for strongly correlated
fractional-QH edges.\cite{uz:prb:99}  The fact that
Eq.~(\ref{allowed}) ensures Fermi statistics has been
regarded\cite{leewen} as strong support for this line of
argument. However, the numerical results discussed in
Sec.~\ref{num} suggest that there is no basis for this
requirement. Assuming that a simple bosonization identity of the
form given in Eq.~(\ref{simpbos}) holds, we can extract the
values $\xi_0^2\approx 0.51\pm 0.02$ and $\xi_1^2\approx 0.65\pm
0.06$ from the data for matrix elements of anticommutators
$\{\tilde\psi_{0}^\dagger,\,\tilde\psi_{0}^\dagger\}$ and
$\{\tilde\psi_{1}^\dagger,\,\tilde\psi_{1}^\dagger\}$. (See the
Appendix for details of the calculation.) The value of $\xi_1^2$
deviates significantly from that expected within $\chi$LL theory
($1/2$) because $\tilde\psi_{1}^\dagger$ does not satisfy Fermi
statistics. While the standard deviations from the average
extracted $\xi_{0,1}^2$ are reasonably small, we have to caution
the reader by noting that the assumption of a simple bosonization
identity would imply symmetry of $B_{n,n^\prime}/A_{n,n^\prime}$
under exchange $n\leftrightarrow n^\prime$. Clearly, no such
symmetry is exhibited in our data. The significantly large
deviation between $B_{0,1}/A_{0,1}$ and $B_{1,0}/A_{1,0}$ raises
a big question mark: unless rather complex features are ascribed
to the Klein factors, a consistent interpretation of our data
using a local bosonization formula like Eq.~(\ref{simpbos}) is
impossible.

\section{Discussion and Conclusions}
\label{concl}

We have shown, by explicit numerical calculation of
anticommutator matrix elements, that the projection of the
lowest-Landau-level electron creation operator onto the
low-energy edge-excitation Fock subspace of a $\nu=2/5$
incompressible quantum Hall state does {\em not\/} satisfy Fermi
statistics. We observe a consistent dependence of the 
anticommutation rules on the particular procedure for adding
electrons, {\it i.e.}, on the change in quasiparticle number that
accompanies particle addition. We find that the numerical data
cannot be consistently interpreted by assuming any simple
generalization of conventional bosonization identities. In
particular, the expression for the electron operator solely in
terms of rigid edge deformations (magnetoplasmon modes), which
two of us argued for previously\cite{uz:prb:99} on heuristic
grounds is also not supported by our calculations. Since our
present study was performed for a system with short-range
interactions, however, we cannot exclude the possibility that
real QH samples where long-range Coulomb interactions are present
may be consistently described by such a bosonization identity, as
is suggested by the amazing experimental finding that the
tunneling-$IV$ exponent $\alpha \approx 1/\nu$.

Our numerical study demonstrates that the specific form of the
boson representation of the edge electron creation operator
cannot be inferred by postulating Fermi statistics of the
projected edge-electron operator. Alternatives\cite{uz:prb:99} to
the $\chi$LL expressions\cite{wen:int:92} cannot be ruled out on
these general grounds. However, our present data for the
short-range-interacting case supports neither the conventional
chiral-Luttinger-liquid picture nor any simple alternative.
This points strongly towards the possibility that there is no
simple universal {\em local\/} bosonization identity for the
edge-electron operator, a conclusion also reached in an
independent recent study.\cite{jainpreprint} If true, this likely
implies that electronic properties of fractional-quantum-Hall
edges depend crucially on sample specifics.

\begin{acknowledgments}
Useful and stimulating discussions with A.~Auerbach,
J.~T.~Chalker, H.~Fertig, F.~D.~M.~Haldane, B.~I.~Halperin,
J.~Jain, R.~Morf, N.~Read, E.~Shimshoni, S.~H.~Simon, and
D.~J.~Thouless are gratefully acknowledged. J.J.P.\ acknowledges
support from the Spanish CICYT (Grant No.~1FD97-1358) and the
Generalitat Valenciana (Grant No.~GV00-151-01). U.Z. thanks the
Aspen Center for Physics for hospitality during the 2000 Summer
Workshop on Low-Dimensional Systems and the German Science
Foundation (DFG) for partial support through Grant No.~ZU~116/1.
A.H.M.\ was supported by the National Science Foundation under
grant DMR~0115947. 
\end{acknowledgments}

\appendix*
\section{Anticommutators calculated from bosonization}

Within the bosonization formalism, the quantities
$A_{n,n^\prime}$ and $B_{n,n^\prime}$, defined by
Eqs.~(\ref{define}), do not depend on $N$ or $N_{\text{qp}}$. We
find $A_{n,n^\prime}=z$ and
\begin{widetext}
\begin{equation}\label{BAratio}
\frac{B_{n,n^\prime}}{A_{n,n^\prime}} = \chi_{n,n^\prime}
\sum_{\{\eta\}}\bra{N+1,
N_{\text{qp}}+n^\prime,0}e^{-i\phi_n(0)}\ket{N+1,N_{\text{qp}}+
n^\prime,\{\eta\}}\bra{N+1,N_{\text{qp}}+n^\prime,\{\eta\}}e^{-i
\phi_{n^\prime}^\dagger(0)}\ket{N+1,N_{\text{qp}}+n^\prime,0}\, ,
\end{equation}
%\end{widetext}
where $\ket{N+1,N_{\text{qp}}+n^\prime,\{\eta\}}$ are excited
states with excess (boson) momentum equal to $\Delta
K_{\text{F}}^{(n,n^\prime)}$. Contributions from Klein factors
are contained in $\chi_{n,n^\prime}$ which satisfies $\chi_{n,n}=
1$. To better understand the two-branch case, it is useful to
first consider the situation when only a single branch of
edge excitations is present.

\subsection{Single-branch case}

At the edge of a QH sample at filling factor equal to $1/(2p+1)$,
a single branch of edge excitations exists which corresponds to
the edge-magnetoplasmon (charged) mode. [Within the formalism
described above, we would have to set $\xi_n\equiv 0$, $\phi_n=
\phi^{\text{(c)}}\equiv\phi$, and let $\nu\to 1/(2p+1)$ in order
to describe the single-branch edge.] Denoting by $a^\dagger_{Q}$
($a_{Q}$) the bosonic creation (annihilation) operator for an
edge-magnetoplasmon excitation with wave number $Q$, we can give
expressions for both the phase field entering the bosonization
identity and the excited states appearing in Eq.~(\ref{BAratio}):
\begin{subequations}
\begin{eqnarray}\label{phidef}
\phi(\theta) &=& \frac{1}{\sqrt{\nu}}\sum_{Q>0}\frac{e^{iQ\theta}
}{\sqrt{Q}}\, a_Q \quad , \\
\ket{N+1, 0, (l)} &=& \prod_r\frac{\big[a_{0,r}^\dagger
\big]^{l_r}}{\sqrt{l_r!}} \ket{N+1, 0, 0}\quad ,
\end{eqnarray}
\end{subequations}
with $\sum_r r l_r = \Delta K_{\text{F}}=2p+1$. Here, $r>0$ and
$l_r\ge 0$ are integers, and $(l)$ is a {\em partition\/} of the
integer $\Delta K_{\text{F}}$ labeling possible many-body excited
states. Straightforward calculation yields
%\begin{widetext}
\begin{equation}
\bra{N+1,0,0}e^{-i\phi(0)}\ket{N+1,0,(l)}=\bra{N+1,0,(l)}
e^{-i\phi^\dagger(0)}\ket{N+1,0,0} = \prod_r \left[
(-\nu\,r)^{l_r}\, l_r!\right]^{-\half}\quad .
\end{equation}
%\end{widetext}
Inserting this result into Eq.~(\ref{BAratio}), we find
\begin{subequations}
\begin{eqnarray}
\frac{B}{A}&=&\sum_{(l)} \prod_{r\in (l)} \left[(-\nu\,r)^{l_r}\,
l_r!\right]^{-1}\quad ,\\
&=&\frac{(-1)^{\Delta K_{\text{F}}}}{\Delta K_{\text{F}}!}
\sum_{s=0}^{\Delta K_{\text{F}}}\left(\frac{1}{\nu}\right)^s 
\sum_{(l)_s} {\mathcal S}^{(s)}_{\Delta K_{\text{F}}}\quad ,
\end{eqnarray}
\end{subequations}
where $(l)_s$ denotes a partition with $s$ cycles, and ${\mathcal
S}^{(s)}_{\Delta K_{\text{F}}}$ denotes Stirling numbers of the
first kind.\cite{abramowitz} (Klein factors are irrelevant for
calculating $B/A$ of a single-branch edge.) Using relation 24.1.3
from Ref.~\onlinecite{abramowitz}, and introducing the
generalized binomial coefficients,
$$\left(\begin{array}{c}a\\n\end{array}\right)=\frac{\Gamma(a+1)}
{\Gamma(a+1-n) n!}\quad ,$$
we find
\begin{equation}\label{singres}
\frac{B}{A} = (-1)^{\Delta K_{\text{F}}}\,\, \left(
\begin{array}{c}\nu^{-1}\\\Delta K_{\text{F}}\end{array}
\right)\quad .
\end{equation}
Obviously, because $\nu^{-1}\equiv\Delta K_{\text{F}}=2p+1$, we
find that Fermi statistic holds for the bosonized expression of
the projected edge--electron operator at Laughlin--series
filling factors: $B/A=-1$.

\subsection{Two-branch case}

Lets now consider again the case $\nu=2/(4p+1)$ where two edge
branches with the same chirality are present. Many-body excited
states entering Eq.~(\ref{BAratio}) are then the eigenstates of
the quadratic edge-density-wave Hamiltonian which are spanned by
bosonic operators that are, in general, some orthogonal linear
combinations of the charged and neutral-mode excitations. The
coefficients in the linear transformation that relate the charged
and neutral modes to the bosonic normal modes of the Hamiltonian
depend on microscopic details, {\it i.e.}, the velocities of the
charged and neutral modes as well as their coupling via
interactions. However, it turns out that we do not need these
coefficients explicitly. It suffices to know that the phase
fields entering the bosonization formula can be expressed in
terms of the normal modes, $\phi_n=\sum_{j=1}^2 \eta_j^{(n)}\,
\varphi_j$, where
\begin{subequations}
\begin{eqnarray}
\eta_1^{(n)}&=&\frac{\cos\alpha}{\sqrt{\nu}}-\xi_n\,\sin\alpha
\quad , \\
\eta_2^{(n)}&=&\frac{\sin\alpha}{\sqrt{\nu}}+\xi_n\,\cos\alpha
\quad .
\end{eqnarray}
\end{subequations}
The fields $\varphi_j$ are defined in terms of the bosonic
operators $a_{j,Q}$ that annihilate excitations of the $j$th
normal mode in analogy to Eq.~(\ref{phidef}), and microscopic
details of the edge determine the value of $\alpha$. Excited
states entering Eq.~(\ref{BAratio}) are direct products of
excited states in the two normal-mode subspaces whose combined
excess momenta equals $\Delta K^{(n,n^\prime)}_{\text{F}}$.
Within each of the normal-mode sectors, total excess momentum is
partitioned among the possible many-boson states as in the
single-branch case. We can, therefore, employ the
result~(\ref{singres}) and find
%\begin{widetext}
\begin{subequations}
\begin{eqnarray}
\frac{B_{n,n^\prime}}{A_{n,n^\prime}} &=& \chi_{n,n^\prime}\,\,
(-1)^{\Delta K^{(n,n^\prime)}_{\text{F}}}\,\sum_Q \left(
\begin{array}{c}\eta_1^{(n)}\eta_1^{(n^\prime)}\\Q\end{array}
\right)\,\left(\begin{array}{c}\eta_2^{(n)}\eta_2^{(n^\prime)}\\
\Delta K^{(n,n^\prime)}_{\text{F}}-Q \end{array}\right)\quad , \\
&=&\chi_{n,n^\prime}\,\, (-1)^{\Delta
K^{(n,n^\prime)}_{\text{F}}}\,\left(\begin{array}{c}\eta_1^{(n)}
\eta_1^{(n^\prime)}+\eta_2^{(n)}\eta_2^{(n^\prime)}\\
\Delta K^{(n,n^\prime)}_{\text{F}}\end{array}\right)\quad .
\end{eqnarray}
\end{subequations}
\end{widetext}
Here we have used the addition theorem for generalized binomial
coefficients. (See, e.g., Section 12.2 of
Ref.~\onlinecite{marmur}.) Straightforward inspection shows that
the sum $\sum_j\eta_j^{(n)}\eta_j^{(n^\prime)}=\nu^{-1} + \xi_n
\xi_{n^\prime}$ is universal, {\it i.e.}, independent of $\alpha$.
Using that, we can write
\begin{equation}\label{twores}
\frac{B_{n,n^\prime}}{A_{n,n^\prime}} = \chi_{n,n^\prime}\,\,
(-1)^{\Delta K^{(n,n^\prime)}_{\text{F}}}\,\left(\begin{array}{c}
\nu^{-1} + \xi_n\xi_{n^\prime}\\
\Delta K^{(n,n^\prime)}_{\text{F}}\end{array}\right)\quad .
\end{equation}

With the choice of values for $\xi_n$ according to $\chi$LL
theory, which is given in Eq.~(\ref{allowed}), one finds
$\nu^{-1} + \xi_n\xi_{n^\prime}\equiv\Delta K^{(n,
n^\prime)}_{\text{F}}$. Since $\Delta K^{(n,n)}_{\text{F}}$ is
always odd, and with a suitable definition of Klein
factors\cite{vondelft} such that $\chi_{n,n^\prime}=(-1)^{n+
n^\prime}$, one obtains $B/A=-1$.

Two of the authors suggested previously~\cite{uz:prb:99} the
choice $\xi_n=0$. For the special case of $p=1$, our result
(\ref{twores}) would predict $B_{0,0}/A_{0,0}=B_{1,1}/A_{1,1}=-5/
16$. Additional assumptions about Klein factors are then
necessary for a consistent description of $B_{0,1}/A_{0,1}$ and
$B_{1,0}/A_{1,0}$.

Having obtained data for $B_{n,n^\prime}/A_{n,n^\prime}$, from
numerical calculations, we used Eq.~(\ref{twores}) to extract the
values of $\xi_n$ and check whether a consistent description
using a simple bosonization identity of the form~(\ref{simpbos})
is possible.

%\bibliography{general,qhegen,qhedge,1deg,myself}

\begin{thebibliography}{40}
\bibitem[{\citenamefont{Palacios}(2002)}]{jjp:note}
\bibnamefont{Similar numerical calculations find evidence for the
failure of Fermi statistics at single-branch QH edges when the
Coulomb interaction is considered instead of the
hard-core-interaction model. We find $B/A \approx -0.98$ for
small ($N=6$ and $N=7$) finite-size $\nu=1/3$ QH droplets.
However, we cannot, at this point, draw any conclusions regarding
how these discrepancies evolve as the thermodynamic limit is
approached. Numerical evidence, based on Coulomb-interaction
model calculations, for non-universality at $\nu=1/3$ edges was
also obtained in
Refs.~\onlinecite{gold:prl:01,gold:prb:01,jain:ssc:01,kun:prl:02}.}
\end{thebibliography}

\end{document}